\documentclass[aps, twocolumn, superscriptaddress, showpacs]{revtex4}

\usepackage[T1]{fontenc}
\usepackage[pdftex]{graphicx}
\usepackage{amsmath}
\usepackage[utf8]{inputenc}

\begin{document}

\title{Blocking of phase transitions in liquid crystal 5*CB (isopentylcyanobiphenyl) as a result of surface interactions at the nano-membranes}

\author{A. Bąk}
\email{abak@prz.edu.pl}
%\homepage[]{Your web page}
%\thanks{}
%\altaffiliation{}
\affiliation{Rzeszów University of Technology, W. Pola 2, 35-959 Rzeszów, Poland}

\author{K. Chłędowska}
%\email[kch@prz.edu.pl]{Your e-mail address}
%\homepage[]{Your web page}
%\thanks{}
\affiliation{Rzeszów University of Technology, W. Pola 2, 35-959 Rzeszów, Poland}

\author{P. Inglot}
\affiliation{Rzeszów University of Technology, W. Pola 2, 35-959 Rzeszów, Poland}

\begin{abstract}
In this paper we present results of dielectric measurements of the liquid crystal (LC) 5*CB arranged in the porous matrices with a pore diameter of 100 and 20nm. We analyze the effect of surface interactions on the dynamics of molecules. The results were compared with the results of the bulk 5*CB. The most important result is the blocking of phase transition of 5*CB into the solid phase in a matrix of 20 nm.
\end{abstract}

\date{\today }
%\pacs{61.30.−v, 64.70.pp, 68.08.−p, 77.22.Gm}

\maketitle

{\it Introduction.}
Liquid crystal displays (LCD) are usually used in a variety of devices (TV, computer screens, mobile, watches, etc.). Improving the quality of the displayed image forces the construction of new displays with more densely packed pixels. Miniaturization reduces the amount of material used in the construction of the display, but properties of LCs confined in a small cell differ from the properties of bulk LCs ~\cite{{1},{1-1},{2}}. Moreover, displays designed to display dynamic images must be characterized by the lowest possible inertia. Pixels are sequentially on/off at each frame, it makes the time required to enable/disable each of them must be increasingly shorter (in fast displays). This time is limited by the time required on reorientation liquid crystal molecules in an electric field. In each LCD liquid crystal is placed in the cell, and therefore a significant impact on the ability to reorientation of the molecules play their interactions with the cell surface~\cite{3}.

In this paper we present the influence of surface effects on the dynamics of the liquid crystal molecules of the chiral 5*CB depending on the size of the cell in which this material was placed. Our results for 5*CB confined in 100~nm matrix were shown in~\cite {4}. Now, we present the new results for 5*CB confined in matrix 20~nm and we compare them with the results obtained for the bulk material and the 5*CB confined in matrix 100~nm.

\begin{figure}
\includegraphics[width=0.45\textwidth]{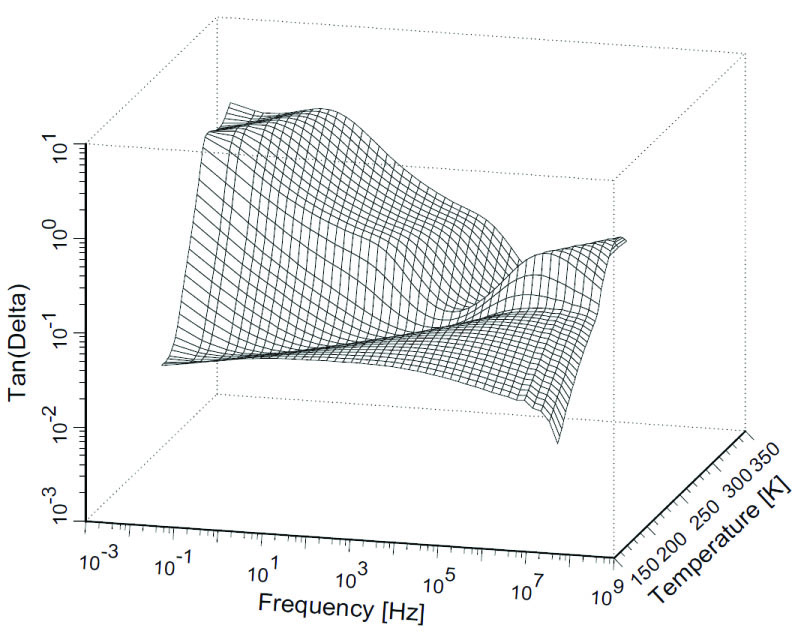}
\includegraphics[width=0.45\textwidth]{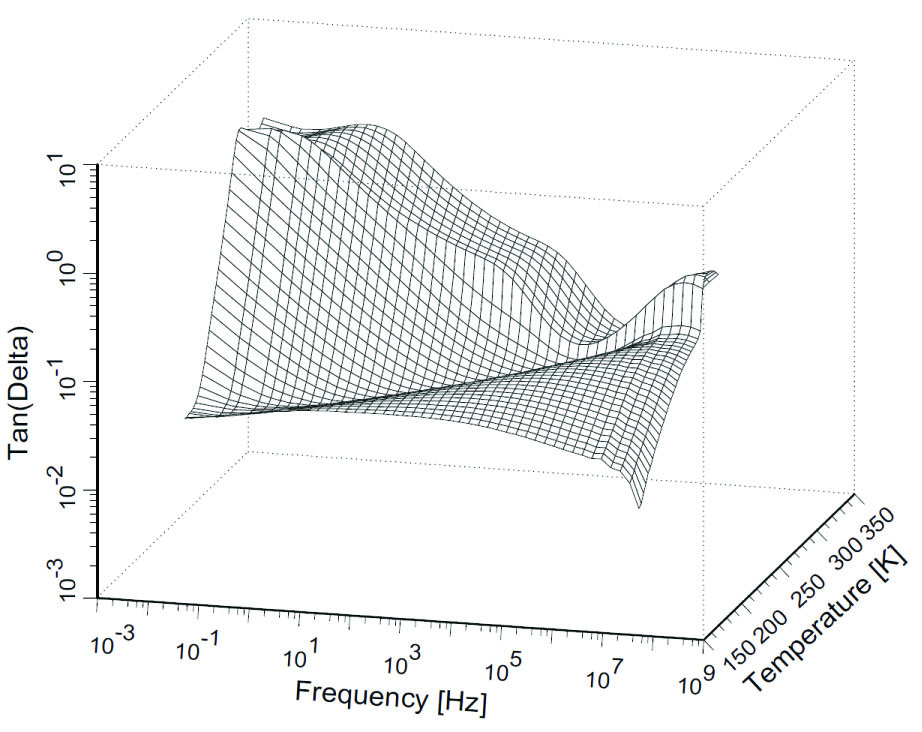}
\caption{\label{fig1}Dependence $tan\delta$ on temperature and frequency of 5*CB confined in matrix of pore size 100 nm during cooling (top) and heating (bottom)}
\end{figure}

{\it Experiment.}
The dynamics of the LC 5*CB molecules have been studied by broadband dielectric spectroscopy (BDS), using the Novocontrol Concept 80 system. A detailed description of the measuring apparatus and method of sample preparation has been described in the works ~\cite{{4}}. Dielectric loss tangent measurements were made as function of the frequency of the electric field in the range of $\rm 10^{-2} - 10^{7}$ Hz, and at temperature in the range of 180 - 310 K. As measurement cells we used a porous matrices (Anopory, Whatman Co.). They are porous blotting paper with a thickness of $\rm 60 \mu$m, having a well-defined size and shape of the pores. They have the shape of the tubes and they are perpendicular to the electrode's surface. For the measurements we used 5*CB, synthesized at the Military University of Technology Warsaw, confined in a porous matrices with pore size of 100 nm (sample I) and 20 nm (sample II).

{\it Results and discussion.}
Fig. 1 presented dielectric loss tangent dependence on frequency and temperature obtained during cooling and heating of the sample I. The dynamics of molecules 5*CB is complex. Four clearly separated peaks in the curves tan$\delta$(f, T) indicate the existence of four relaxation processes, while in the bulk of 5*CB are only two dynamic processes ~\cite{5}. Increasing the number of dynamic processes is caused by the interaction of LC molecules with the walls of the pores ~\cite{{4},{6},{7}}. Most significant difference is dramatic reduction in the amplitude of the all processes of relaxation during cooling at a temperature about 260K. It indicates that there is phase transition during which occurs inhibition of the reorientation of the entire molecules or parts thereof. This is not isotropic-cholesteric phase transition observed in bulk 5*CB, which is related to a slight decrease in amplitude ($\Delta\varepsilon' \approx$ 1 ~\cite{5}). Most likely it is transition to a metastable solid phase, in which is going through a dynamic process with a large distribution of relaxation times and small amplitude. This phase is melting in temperature about 282K and it is confirmation that it is metastable phase ~\cite{{8},{8-1}}.

Fig. 2 shows the relationship tan$\delta$(f, T) during the cooling and heating of the sample II. The results indicate the existence of at least three relaxation processes.

First high frequency process is associated with the rotations of molecules around short axes, and is called as alpha relaxation. The next two overlapping relaxation processes are observed in the low frequencies. Complexity of the low-frequency process clearly indicate differences in the spectrum of the cooling and heating of the sample II in the temperature range 260-280 K. During the sample cooling, one of them is inhibited, therefore during heating does not appear in the observed spectrum. Also, the values $\rm \varepsilon'$ determined at the frequency $\rm f = 10^{-1}$~[Hz] in this temperature range is significantly smaller on heating than during cooling (Fig. 3a)

\begin{figure}
\includegraphics[width=0.45\textwidth]{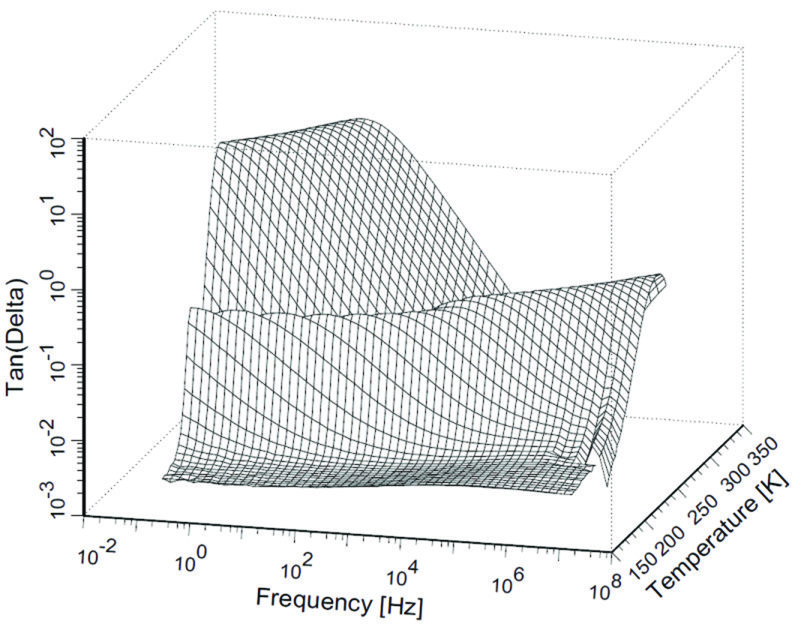}
\includegraphics[width=0.45\textwidth]{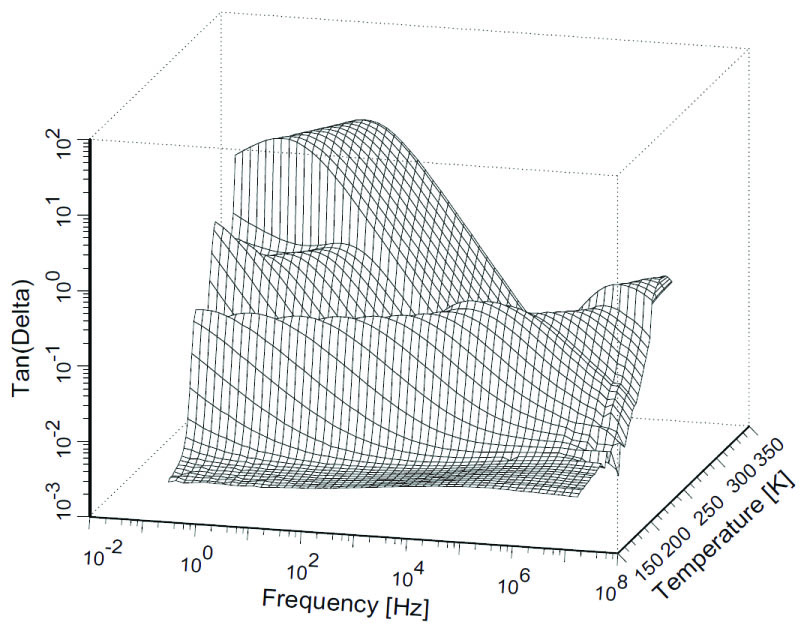}
\caption{\label{fig2}Dependence $tan\delta$ on temperature and frequency of 5*CB confined in matrix of pore size 20 nm during cooling (top) and heating (bottom)}
\end{figure}

The results obtained during the cooling of the sample II are very similar to those obtained for the bulk of 5*CB ~\cite{5}. During cooling in the whole temperature range are observed relaxation alpha. This indicates that, unlike for sample I, and similarly as in bulk material, does not observe crystallization into the metastable solid phase. Since the pore size of the sample II is much less than for sample I, we believe that significant part of molecules  is anchored on the surface of the pores. Thus formed structure where the ability for  rotations of molecules (or flip-flop  motions) are not significantly limited, but due to the blocking of the translation of the molecules, does not occur the transformation into metastable solid phase. We believe that the sample may form a structure resembling to an amorphous glassy phase generated not by freezing the sample, but with decreasing temperature and freezing the structure by anchoring molecules on the pore walls.

\begin{figure}
\includegraphics[width=0.45\textwidth]{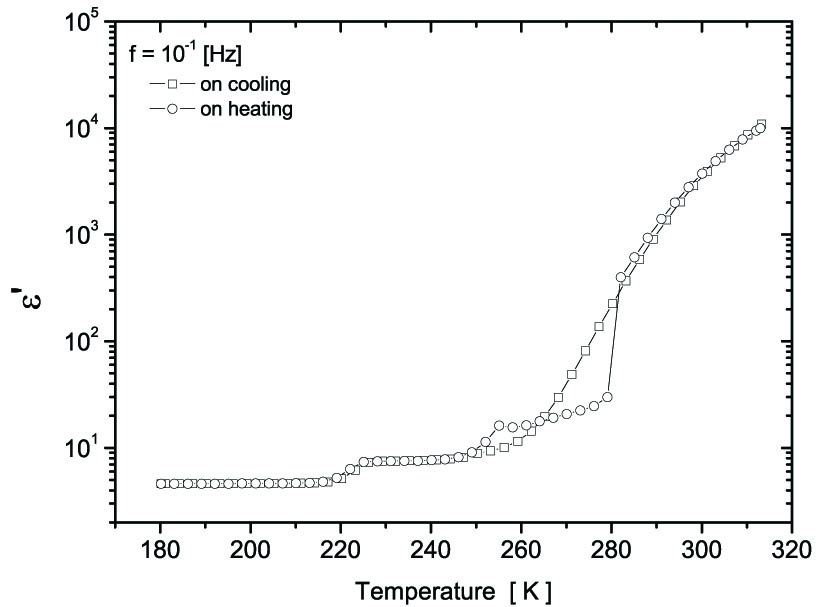}
\includegraphics[width=0.45\textwidth]{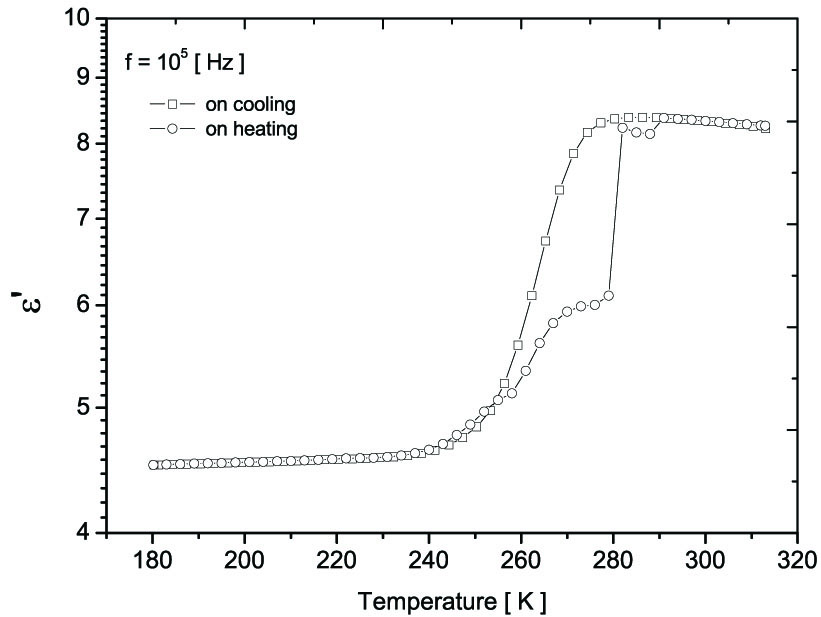}
\caption{\label{fig3b}Real permittivity $\varepsilon'$ vs T for frequencies $\rm 10^{-1}$Hz (top) and $\rm 10^{5}$Hz (bottom) for 5*CB confined in matrix of pore size 20 nm during cooling and heating }
\end{figure}

Main difference compared to the bulk 5*CB is visible during heating of the sample II. In bulk 5*CB, during cooling, isotropic liquid transforms into chiral nematic and then formed glassy phase. During heating glassy phase occurs of a gradual unfreezing, resulting in increased freedom of rotations and translation of the molecules. At temperature about 240 K there is sudden inhibiting of rotation of molecules. Depending on the applied thermal treatment can then be created metastable or, under special conditions, the stable solid phase ~\cite{8}. In the solid phase the value of $\rm \varepsilon'$ decreases to a value of $\rm \varepsilon'$ in the glass.

For the sample II, during heating, alpha relaxation is not blocked and it is observed in the whole temperature range. Just as for the low frequency relaxation process, some slight reduction in the amplitude of the dielectric loss is observed in the temperature range 260 - 280 K. Therefore, we can formulate very important conclusion, that during heating of the sample II, transition to metastable solid phase, observed both in the sample I and bulk material, is blocked. Reducing the $tan\delta$ (Fig. 1 bottom) and $\varepsilon'$ (Fig. 3 bottom) indicates on a reduction in the effective component of the dipole moment in the direction of the external electric field, or freezing of the motions certain part of rotating molecules. Relative change in the amplitude dielectric losses for low- and high-frequency processes in the temperature range of 260 - 280 K is almost the same (approximately 20\%) and it indicating, that being blocked both slow and fast reorienting processes of the same molecules.

{\it Conclusion.}
Both the results obtained for samples I and II differ significantly from the results of bulk material. The main differences are related to (i) changes in the sequence of phase transitions of the sample I during cooling, (ii) changes in the sequence of phase transitions for sample II upon heating, (iii) increasing the number of relaxation processes for 5*CB confined in nanopores matrices.

For 5*CB confined in a matrix of 100 nm (sample I), the most significant difference is observed during cooling, when the phase transition into the metastable solid phase occurs (not observed in bulk material).   
The shape of nanopores and their axial orientation relative to the external electric field, forces the axial arrangement of molecules within the pores, making it impossible to formation of cholesterol phase during the cooling. Axial arrangement of molecules promotes the formation of metastable solid phase ~\cite {4}.

Sample II during cooling behaves the same as the bulk material. A significant difference occurs during the heating, after earlier cooling. In this case, the phase transition to metastable solid is blocked, which was always observed for the bulk material with the same thermal treatment.
We think that the small size of the pores hinders translational motions of molecules, which additionaly can be anchored on the pore walls. This produces a pseudo-amorphous structure, which is temperature-stable. This phase does not transform into metastable solid phase, either during the cooling (such as the sample I) or upon heating (such as in bulk material).

\end{document}